\documentstyle[12pt]{article}
\begin{document}

\title{Radiation in a Fractal Cosmology}

\author{A. K. Mittal\thanks{\small Department of Physics, 
University of Allahabad, Allahabad - 211 002, India; 
mittal\_a@vsnl.com}, Daksh Lohiya\thanks{\small 
Department of Physics and Astrophysics, University of Delhi, 
New Delhi--7, India; dlohiya@ducos.ernet.in}}

\date{}
\maketitle 

\vspace{-1.2cm}
\begin{center}
\em Inter University Centre for  Astronomy and Astrophysics,
Postbag 4, Ganeshkhind, Pune 411 007, India
\end{center}

PACS. 98.80-k - Cosmology

PACS. 98.65Dx -	Superclusters; large-scale structure of the Universe

PACS. 05.45Df - Fractals

\begin{abstract}

	It is shown that Homogeneous radiation can not be included in the Fractal Cosmological
model obtained earlier by assuming an isotropic fractal cosmography, 
General Relativity and the Copernican Principle.

\end{abstract}

\pagebreak
\section*{} 
  
  $~~~~$Recently a fractal cosmological model has been proposed \cite{akmf}
  that follows naturally if one assumes (i) that matter in the Universe is 
  distributed like an isotropic fractal of dimension 2, (ii) the General Theory
  of Relativity and (iii) the Conditional Cosmological Principle \cite{mandl}. 
  Hitherto, the absence of a physical model has been an impediment for subjecting
fractality to precision cosmological tests \cite{peebls,piet}.
  
  In a fractal Universe, density is not defined at any point. The concept of 
  density has to be replaced by that of a ``mass measure'' defined over sets. 
  The mass measure as obtained by any observer moving with the cosmological fluid
  (part of the fractal) will be the same as that obtained by any other observer.
  However, observers in a region of void are precluded, because spheres drawn with such 
  points (not belonging to the fractal) as centre, will be empty with probability 1
  \cite{chaosbook}.
  
  It was shown in \cite{akmf} that the time--time component $G^{00}$ of the Einstein tensor 
  for any observer on an occupied point of the fractal (galaxy), must be of the form:
$$
G^{00}_{\rm fractal}(t,\chi,\theta,\varphi)  = f(\chi ) G^{00}_{\rm FRW}(t) \eqno{(1)}
$$
where $G^{00}_{\rm fractal}$ and $f(\chi)$ are to be regarded as ansatz to compute 
measures to satisfy the integrated Einstein's equations over any 3-volume in the 
constant time hypersurface and:
$$
G^{00}_{\rm FRW}  =  3\left\{\left({{\dot a}\over a}\right)^2 + 
{k\over {a^2}}\right\}\eqno{(2)}
$$
is the $G^{00}$ obtained from the FRW metric:
$$
ds^2 = - dt^2 + a^2(t)\{d\chi^2 + \Sigma^2(d\theta^2 + \sin^2{\theta} d\varphi^2)\} \eqno{(3)}
$$
where
$$
\Sigma \equiv \sin{\chi} \,\,\,\,\,\,\,\,\,\,\,\,\mbox{if $k 
\equiv {K\over{|K|}}= +1$ (positive spatial curvature)}\eqno{(4)}
$$

$$
\Sigma \equiv \chi  \,\,\,\,\,\,\mbox{if $k 
\equiv K = 0$ (zero spatial curvature)}\eqno{(5)}
$$

$$
\Sigma \equiv \sinh{\chi} \,\,\,\,\,\,\,\,\,\,\,\,\mbox{if $k 
\equiv {K\over {|K|}} = -1$ 
(negative spatial curvature)}\eqno{(6)}
$$
	Let $S^3_P(R)$ denote a hypersphere of radius $R$ centered at $P$ 
on the hypersurface of constant
time $t$, and assume $M_P(R)$, the mass enclosed in $S^3_P(R)$,
to have  a fractal dimension $D$. Then:
$$
M_P(R) =
\left \{
\begin{array}{lll}
&  C(t) R^D\\
& 0
\end{array}
\right.
\begin{array}{l}
\mbox{if P $\in$ the fractal }\\
\mbox{otherwise} 
\end{array} \eqno{(7)}
$$
The integrated Einstein's eqns. give:
$$
G^{00}_{\rm FRW}(t) = 6 C(t) a^{D-3}(t)\eqno{(8)}
$$
and 
$$
f(\chi ) = {D\over 3}\Sigma^{D-3}(\chi )\eqno{(9)}
$$
We denote the ``fractal mass density '' $C(t)$ by $C_a$, because 
the scale factor $a$ is a function of time $t$. The `fractal mass density '' for
scale factor $a(t)$ is related to that at the present scale factor $a_0$ by:
$$
a^D(t)C_a = a_o^DC_{a_o}\eqno{(10)}
$$
Eqns (2), (8) and (10) yield:
$$
3\left\{\left({{\dot a}\over a}\right)^2 + {k\over {a^2}}\right\} 
= 6 {a_o^D\over a^D} C_{a_o}a^{D - 3}\eqno{(11)}
$$
Eqns. (9) and (11)
describe the fractal cosmology completely for a universe in which matter is 
distributed as a fractal of dimension 2 and ``fractal mass density'' $C_{a_o}$
when the scale factor is $a_o$.

	There is evidence to suggest \cite{lab} that matter distribution 
in the Universe could be a fractal of dimension 2. However, there is no
evidence to suggest that radiation is not homogeneous to a very good
accuracy.

	So let us try to include homogeneous radiation in our model described by radiation 
density $\rho_\gamma$ which scales with scale factor $a(t)$ as:
$$
\rho_\gamma a^4 = \rho_{\gamma_0} a_0^4 \eqno{(12)}
$$
Then the integrated Einstein's equations will yield:
$$
4\pi G^{00}_{\rm FRW}(t) a^3(t)\int_0^\chi f(\chi )\Sigma^2(\chi ) \Sigma'(\chi )d\chi
$$
$$
= 8\pi \{C(t) a^D(t) \Sigma^D(\chi ) 
+ 4\pi \rho_\gamma (t) a^3(t) \int_0^\chi \Sigma^2(\chi ) \Sigma'(\chi )d\chi\} 
$$
$$
= 8\pi \{ C_{a_0} a_0^D \Sigma^D(\chi ) 
+ 4\pi\rho_{\gamma_0}{a_0^4(t)\over {a(t)}} 
\int_0^\chi \Sigma^2(\chi ) \Sigma'(\chi )d\chi\} \eqno{(13)}
$$
This equation will have solutions only for $D = 3$, that is homogeneous distribution of matter.

	In \cite{joyce}, fractal matter was treated as a perturbation in a radiation 
dominated universe. The fractal cosmology of \cite{akmf}, which follows naturally 
from first principles, incorporates fractal distribution of matter a priori.

	We have shown that homogeneous radiation can not be incorporated in a fractal
model. However, eqn(13) may throw more light on the treatment of fractal matter as a 
perturbation in a radiation dominated universe as done in \cite{joyce}. 
Fractal matter can be neglected if the condition:
$$
C_{a_0} \ll {{4\pi\rho_0a_0^{4-D}}\over {3a(t)}}\Sigma^{3-D} \eqno{(14)}
$$
is satisfied. Clearly for a given
epoch, this condition will be violated for sufficiently small $\chi$. This suggests
a need for incorporating a lower cutoff in the fractal distribution of matter, which will 
also make the model more realistic. However, for a given low cut-off, the assumption 
of radiation dominance will fail for sufficiently large $a(t)$. (For open ($k = -1$) models,
this would mean for sufficiently large time).

\vskip 1cm
\section*{Acknowledgements}   
   
We thank Inter University Centre for Astronomy and Astrophysics 
(IUCAA) for hospitality and facilities to carry out this research.

\bibliography{plain}

\end{document}